\documentclass[fleqn,usenatbib]{mnras}

\usepackage{newtxtext,newtxmath}

\usepackage[T1]{fontenc}
\usepackage[table,xcdraw]{xcolor}

\DeclareRobustCommand{\VAN}[3]{#2}
\let\VANthebibliography\thebibliography
\def\thebibliography{\DeclareRobustCommand{\VAN}[3]{##3}\VANthebibliography}

\usepackage{graphicx}	
\usepackage{amsmath}	
\usepackage{makecell}
\usepackage{adjustbox}
\usepackage{subfig}

\newcommand{\z}{\phantom{0}}

\title[Dynamical Modelling of the `Jackpot' Lens]{Two-dimensional kinematics and dynamical modelling of the `Jackpot' gravitational lens from deep MUSE observations}

\author[Turner et al.]{
Hannah C. Turner$^{1}$\thanks{Contact e-mail:\href{mailto:hannah.c.turner@durham.ac.uk}{hannah.c.turner@durham.ac.uk}}
Russell J. Smith$^{1}$ and Thomas E. Collett $^{2}$
\\
$^{1}$Centre for Extragalactic Astronomy, University of Durham, Durham DH1 3LE, United Kingdom\\
$^{2}$Institute of Cosmology and Gravitation, University of Portsmouth, Portsmouth, PO1 3FX, United Kingdom
}

\date{Accepted 2024 January 15. Received 2023 December 21; in original form 2023 July 14}

\pubyear{2022}

\begin{document}
\label{firstpage}
\pagerange{\pageref{firstpage}--\pageref{lastpage}}
\maketitle

\begin{abstract}
We present results from the first spatially resolved kinematic and dynamical modelling analysis of the unique SDSSJ0946+1006 (‘Jackpot’) triple-source lens system, where a single massive foreground $z\,=\,0.222$ galaxy multiple-images three background sources at different redshifts. Deep IFU spectroscopic data were obtained using the MUSE instrument on the VLT, which, compared to previous single-slit observations, provides full azimuthal area coverage, high sensitivity (5 hour integration) and high angular resolution (0.5\,arcsec FWHM). To account for the strong continuum contributions from the $z\,=\,0.609$ source, a multiple-component stellar template fitting technique is adopted to fit to the spectra of both the lens galaxy and the bright lensed background arc simultaneously. Through this, we robustly measure the first and second moments of the two-dimensional stellar kinematics out to about 10\,kpc from the centre of the lens, as well as resolving the inner profile inwards to $\sim$1\,kpc. The two-dimensional kinematic maps show a steep velocity dispersion gradient and a clear rotational component. We constrain the characteristic properties of the stellar and dark matter (DM) mass components with a sufficiently flexible parameterised dynamical model and an imposed lensing mass and find a DM density slope of $\gamma\,=\,1.73\substack{+0.17 \\ -0.26}$, i.e. significantly steeper than an 
unmodified NFW profile ($\gamma\,=\,1$) and consistent with a contracted DM halo. Our fitted models have a lensing-equivalent density slope of $\eta\,=\,0.96\pm0.02$, and thus we confirm most pure lensing results in finding a near isothermal profile for this galaxy.
 
\end{abstract}


\section{Introduction}

The structure of the most massive elliptical galaxies provides a window onto the early history of galaxy formation. The 
currently-popular
``two-phase'' formation history of such galaxies  \citep[e.g.][]{oser10} -- with an early starburst-driven phase and subsequent accretion through mergers -- offers a vital framework for understanding the assembly and evolution of galaxies, and the distribution of stars and dark matter (DM) within them. 

Lacking ordered gas dynamical probes, stellar kinematics represent the main method to study the mass distribution in massive elliptical galaxies \citep[][]{cappellari16}. However, in the absence of detailed observations from high resolution data, analyses at any significant redshift are limited to simple dynamical Jeans type models \citep[e.g.][]{Cappellari08}, as opposed to more general triaxial orbit based models \citep[e.g.][]{Schwarzschild79, vandenbosch08}. Additionally, due to the dependence on luminous mass tracers, of which there is an absence of at large radii, and the sensitivity to many model dependant degeneracies \citep[e.g.][]{bender94, carollo95, gerhard96, romanowsky97, vandenbosch97, weijmans09, oldham2016b}, galaxy dynamics alone do not provide sufficient information to disentangle the stellar and dark mass contributions, and hence place robust constraints on the physics of the DM particle.  

In the rare cases of galaxies that gravitationally lens more distant sources, additional constraints are made available. At its simplest, strong
gravitational lensing provides a robust single estimate of the mass projected
within the Einstein radius \citep[][]{treu10_review}.
With a more sophisticated pixel-based analysis \cite[e.g. following][]{Dye05}, lensing can also probe the slope and shape of the mass density profile in the vicinity of the lensed images, at large radii where dynamical studies are not as sensitive \citep[e.g.][]{ritondale19, shajib21, etherington22}. However, lensing-only studies can be susceptible to degeneracies in the lens modelling and inherently lack sensitivity to the distribution of matter far from the lensed arcs. These degeneracies are independent of those affecting kinematic studies \citep[][]{courteau14}.
 
The unification of mass and structure constraints from kinematic and dynamical modelling with the larger scale mass information from strong lensing allows further insight into the intrinsic properties of galaxies and the nature of DM. Access to a diverse range of spatial scales makes 
combined lensing and dynamical studies powerful
in disentangling the stellar and the DM mass distributions of lens galaxies at their characteristic radii, and thus breaking the degeneracies between these two components \citep[e.g.][]{treu04, koopmans06, barnabe09, Auger10, Treu10, oldham18, shajib21}.
For example, measurements of the Einstein radius from lensing studies can provide constraints on a lens galaxy's total enclosed mass, and by combining stellar dynamics with lensing studies, galaxy-scale strong lensing can provide robust measurements of the stellar IMF \citep[e.g.][]{Treu10, Auger10, Smith15}.
The sensitivity of joint lensing and dynamical studies to different mass scales also gives direct insight into the apparent and surprising near-isothermality of mass in early-type galaxies (ETGs) \citep[e.g.][]{koopmans09, Auger10b, li18}. More pertinently, this ‘bulge-halo conspiracy’ describes how the total mass distribution of ETGs can be described by a power law, but the mass profile of neither the baryonic nor dark matter components can be described by a power law on their own \citep[][]{treu04, treu06, humphrey2010}. For an up-to-date review of galaxy-scale strong lensing, including the application of lensing-plus-dynamics studies, see \citet{shajib22}.

Double-source-plane lenses (DSPLs), or compound lenses, are a rare and valuable type of gravitational lens system, occurring when a single foreground lens galaxy simultaneously multiply-images two background source galaxies at different redshifts. The best studied example of a DSPL to date is the `Jackpot' $z$\,=\,0.222 lens (SDSSJ0946+1006), discovered serendipitously by \citet{Gavazzi08} as part of the Sloan Lens ACS survey \citep[]{Bolton06}. The Jackpot system consists of a bright ring at $z$\,=\,0.609 and a further ring at a greater radius, indicating a more distant second source ($z_{\rm{spec}}\,=\,2.035$) from which constraints on the cosmological parameters can be obtained \citep{Collett14,Smith21}.
A further multiply-imaged source at $z$~$\approx 6$ has been reported by \citet{Collett20}, making Jackpot a triple-source-plane lens system.

The Jackpot system hosts one of only a few cases of a dark substructure detected through lensing perturbations \citep[see][]{Vegetti10}. The substructure is cited as having a mass high enough that one would expect it to host a luminous galaxy, as well as having a surprisingly high central density. The primary lens has also been reported to have a steep density slope by \citet{Minor21}, a claim that sits in contention with that of earlier findings in \citet{Collett14}. The authors note that the peculiar inferred properties of the subhalo could be due to a deviation from the CDM paradigm with respect to the particle physics of DM, such as dark matter self-interactions \citep[see][]{Colin02, Vogelsberger12, zavala19, turner21}, in which case one would expect to detect many more highly concentrated substructures in future surveys.
They also note, however, that the substructure properties could be affected by the lack of generality in their model, stating that a more flexible host galaxy combined with spatially resolved kinematics could provide stronger constraints for the subhalo concentration. 

As a result of its astrophysical significance and applications in addressing open cosmological questions, the lensing properties of the Jackpot have been intensively studied.
In contrast to this, and despite the additional advantages that kinematic data can bring to breaking degeneracies, only relatively limited kinematic measurements exist for the lens system \citep[e.g.][]{auger09, Sonnenfeld12, Spiniello15}. These studies used single-slit observations, such as the 1\,arcsec-wide slit used in \citet{Sonnenfeld12}, yielding measurements to $\sim1$\,arcsec from the centre of the lens, and the $2\times1.5$\,arcsec slit of \citet{Spiniello15}, and thus cover only a moderately small radial distance. As a result, this limits the area from which measurements can be obtained to within the bright Einstein ring, effectively excluding the radial ranges at which DM contributions become more significant and resulting in restricted spatial information.

In this work, we present a combined kinematic and dynamical analysis of 
the Jackpot lens galaxy. We apply template fitting methods to deep integral field unit (IFU) spectroscopic data from the MUSE instrument, and employ anisotropic dynamical Jeans models that are robustly constrained by the lensing mass at the Einstein radius, to measure the total 2D-projected density profile slope of the lens galaxy. This paper is organised as follows. In Section~\ref{sec:data}, we describe our IFU data and highlight how they differ from and improve on those of previous studies. In Section~\ref{sec:kinematics}, we detail our use of multiple-component template fitting to the observed galaxy spectra, in order to extract the stellar kinematics of the foreground lens galaxy. Section~\ref{sec:modelling} describes a set of kinematic predictions obtained through dynamical modelling of a gNFW + stars model with varying parametrisations. The observed stellar kinematics are then compared with the dynamical model predictions to recover the best-fitting DM density slope and mass, and the total projected logarithmic density slope for a mass profile with our best-fitting parameters. In Section~\ref{sec:discussion}, we discuss our findings and place them in the context of previous studies of the Jackpot lens. We also consider the robustness of our assumptions and the fundamental limitations of our modelling approach. Findings are summarised in Section~\ref{sec:summary}.

Throughout this work we fix the angular diameter distances of the system assuming flat  $\Lambda$CDM cosmological parameters $\Omega_m\,=\,0.307$, $\Omega_\Lambda\,=\,0.693$ and $h_0\,=\,0.6777$ \citep[][]{planck2014}.
At the lens redshift of $z$\,=\,0.222, the angular scale is  3.69\,kpc\,arcsec$^{-1}$.

\section{Data}
\label{sec:data}

Deep integral field unit (IFU) adaptive-optics-assisted spectroscopic data \citep[project ID 0102.A-0950 as described in][]{Collett20, Smith21} were obtained from a 5.2 hour total integration time with the MUSE instrument on the VLT. When compared to previous single-slit observations, this data provides full azimuthal area coverage, high sensitivity and a high angular resolution (0.5\,arcsec FWHM). As a result, we can measure the two-dimensional stellar kinematic properties out to $\sim$~10\,kpc ($\sim$~2.7\,arcsec) from the centre of the lens, as well as resolving the inner profile inwards to $\sim$1\,kpc. Measuring the kinematics out to larger radii with greater precision than previous studies were able to achieve helps to break the degeneracies between the stellar and DM mass components by probing the radii where DM contributions become more significant. Thus, allowing us to span the relevant range for projected mass slope measurements at the Einstein radius that allow us to place our results into the context of previous pure lensing studies.

\begin{figure}
	\includegraphics[width=\columnwidth]{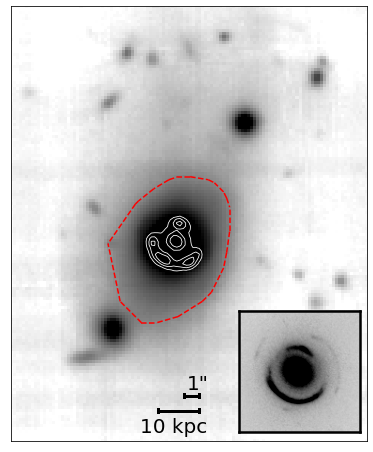}
    \caption{Collapsed MUSE image of the Jackpot triple-source lens system and its environment. The red dashed line denotes the extent of the stellar kinematic measurements, and the white contours highlight the location of the foreground lens galaxy and the first Einstein ring. Inset is a F814W image from HST  of the Jackpot lens and the first and second Einstein rings.}
    \label{fig:data}
\end{figure}

Fig.~\ref{fig:data} shows a collapsed MUSE image of the lens and its environment. Denoted are the extent of the stellar kinematic measurements and the locations of the foreground lens galaxy and the first Einstein ring. Some of the structure visible at low surface brightness is caused by MUSE sensitivity variations, but the overall asymmetry, with an extended plume to the north, is reproduced in other imaging, e.g. Fig. 3 of \citet{Sonnenfeld12}.

The physical scale resolution of our observations (1.86\,kpc for a 0.5\,arcsec FWHM at $z\,=\,0.222$) is an order of magnitude coarser than that of the typical dynamical analyses of nearby early-type galaxies. For example, the ATLAS$^{\text{3D}}$ survey \citep[][]{Cappellari11} which, with an angular FWHM of 1.5\,arcsec and galaxy redshift of $z\lesssim0.01$, has a physical spatial resolution typically $\sim$0.15\,kpc. However, the modelling techniques that we use here are also routinely applied to galaxies with comparable physical resolutions to our data. For example, the MaNGA Survey galaxies \citep[][]{law16}, which have a poorer median spatial resolution of 2.54\,arcsec FWHM for a median redshift of $z\,=\,0.037$, leading to a physical FWHM of 1.8\,kpc.

\section{Stellar Kinematics}
\label{sec:kinematics}

In order to construct dynamical models and constrain the 2D-projected total mass profile slope of the lens, robust measurements of the spatially resolved stellar kinematic properties must be obtained. To achieve this, stellar template fitting is employed, and measurements of the first and second order moments of the lens galaxy's 2D stellar kinematics are acquired. To account for the strong continuum contributions from the higher-redshift source, a multiple-component fitting technique must be adopted. The implementation of spatial binning methods is necessary to achieve a high enough signal-to-noise (S/N) ratio for precise kinematic measurements.

\subsection{Kinematic Template Fitting}
\label{sec:kinematic_templates}

Prior to fitting spectral templates to our measured galaxy spectra, we implement the {\sc VorBin} \citet{cappellari03} two-dimensional adaptive spatial binning method. Fundamentally, this works by evaluating pixels that are near to each other and grouping them together in order to achieve an approximately constant S/N ratio. The data described in this work have been binned into 53 bins. 

To measure the kinematics of the lens galaxy, we make use of the stellar template fitting software {\sc ppxf} \citep{Cappellari04, Cappellari17}, which implements penalised pixel-fitting to extract the moments of the line of sight velocity distribution from galaxy spectra. This method works by searching a library of template spectra over a range of metallicities and ages, in this case an ensemble of simple stellar populations (SSPs), to fit to the observed spectra from each bin and thus take kinematic measurements. With this approach, we consider a wavelength range of 4699\,\r{A} to 7408\,\r{A} and apply the penalised pixel-fitting method to each bin individually. The template library is taken from the E-MILES stellar population models \citep[][]{vazdekis16}, chosen for their broad spectral range (1680 to 50000\,\r{A}), good resolution (FWHM\,=\,2.5~\r{A} from 3540\,\r{A} to 8950\,\r{A}) and age/metallicity coverage (--1.79\,<\,[M/H]\,<\,+0.26 and ages above 30~Myr). Unlike other SSP models that do not extend far enough in to the blue end of the spectrum for us to recover the relatively distant z\,=\,0.609 source, this broad spectral range allows us to fit to the younger spectral features of the source galaxy, as well as the older features of the lens elliptical galaxy. This basic approach works well on the very central regions of the system where there exists little contamination from the source light.

In order to place robust constraints on the mass model, it is desirable to map the kinematic measurements in a range of radii that span the Einstein radius, given our mass constraint at this distance and the high area coverage made available by our MUSE data. However, there are a number of bins in these outer regions where strong continuum contributions from both the source and lens galaxies are present. This demands more complexity in our template fitting than the standard {\sc ppxf} treatment provides and makes it necessary to fit to both lens and source simultaneously, as whilst the source emission lines can be simply masked out, the bright source continuum can not sufficiently be masked without a significant loss of spectra. 

\begin{figure}
	\includegraphics[width=\columnwidth]{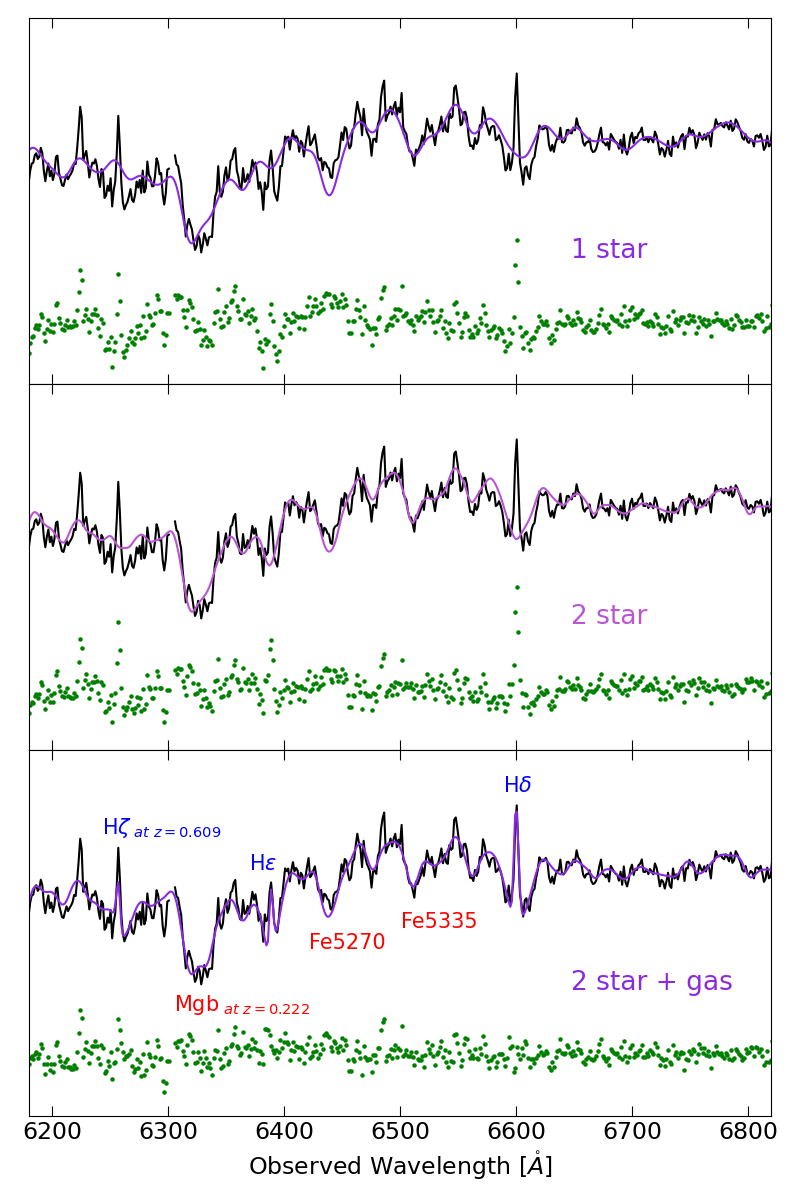}
    \caption{A zoomed-in view of the total fit to the galaxy spectrum for the 1 star, 2 star and 2 star + gas template fitting methods. 
    The 1 star fit highlights how the standard usage of {\sc ppxf} gives a poor fit to the composite spectra, i.e. pixels contaminated by light from the Einstein ring. We can see strong emission lines in the observed data that are not being matched by the total fit template derived from the 2 star fitting method. The 2 star + gas panel demonstrates the significantly improved fit of the total model to the observed spectrum. Also indicated are the absorption features at the lens redshift, and the emission peaks at the source redshift. }
    \label{fig:zoomed_process}
\end{figure}

This is demonstrated in the first panel of Fig.~\ref{fig:zoomed_process}, which shows strong source emission lines and prominent Balmer absorption lines in our observed data that are not being fit by the total fit template. Whilst these lines are mostly present in the spectra of the bluer source galaxy, not accounting for them will affect our ability to recover the kinematics of the lens galaxy as {\sc ppxf} otherwise compromises by using templates from an older but higher-sigma population to account for the source. This motivates the necessity for a multiple-component fitting approach, whereby this problem is alleviated through the addition of both a second set of stellar templates to model the source galaxy, and a set of gas emission templates. The improvement of the total fit to the observed spectra obtained through the addition of these additional components can be seen in the second and third panel of Fig.~\ref{fig:zoomed_process} respectively, and is described in Section~\ref{sec:multi}.

\subsection{Multiple Component Fitting}
\label{sec:multi}

\begin{figure*}
	\includegraphics[width=1.5\columnwidth]{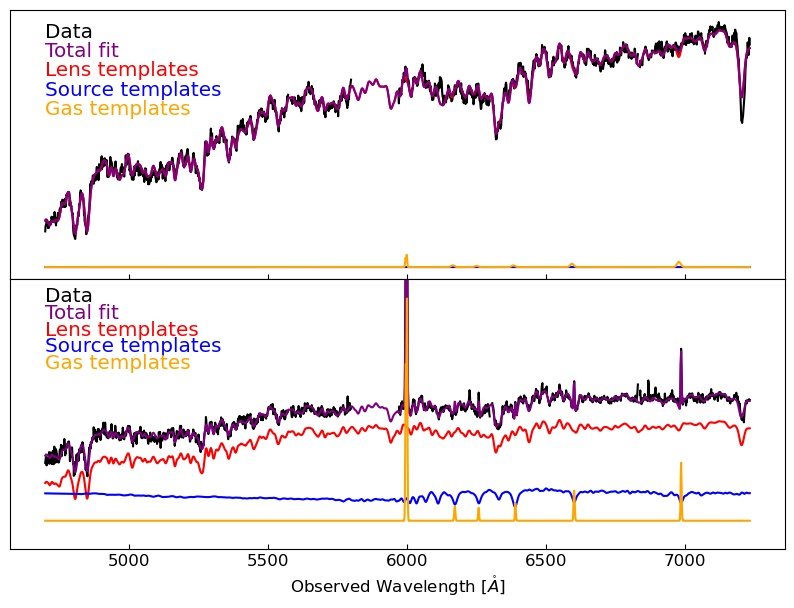}
    \caption{A demonstration of the multiple components used in the total fit to the galaxy spectrum, as demonstrated for both a central bin almost entirely dominated by lens light and a bin near the Einstein radius that is highly contaminated from the lensed source. The stellar kinematic components correspond to the lens galaxy at $z$\,=\,0.222 and the source galaxy at $z$\,=\,0.609. Also present are the two gas components accounting for the [O$_\mathrm{II}$] doublet and Balmer emission lines.}
    \label{fig:how_works}
\end{figure*}

 Fig.~\ref{fig:how_works} shows examples of both a central bin almost entirely dominated by lens light and a highly contaminated bin at the Einstein radius. To each bin spectrum, we mask out prominent subtraction residuals from the brightest sky lines and fit stellar kinematic components corresponding to the lens and the source galaxies and also a small number of gas components. To do this we define a total of four components to be fit:
 
 \begin{itemize}
  \item a combination of SSP templates at redshift $z\approx0.222$ to represent the lens;
  \item a combination of SSP templates at redshift $z\approx0.609$ to represent the bright source;
  \item a gas template at the source redshift, corresponding to the Balmer series emission lines from H$\beta$ to H$\eta$ and with fixed Case-B recombination flux ratios;
  \item a further gas template at redshift z\,=\,0.609 for the [O\,{\sc ii}] doublet\footnote{We note here that the velocity of [O\,{\sc ii}] is not explicitly tied to that of the Balmer lines or the stars.}.
\end{itemize}
 
 Despite the main focus of this fitting being the extraction of the lens kinematics, each component has its own velocity and velocity dispersion in order to recover the lens component without bias\footnote{Recall that the data were binned according to the continuum S/N ratio to optimise the recovery of the lens stellar kinematics, and therefore the binning is poorly configured for spatially resolving the source kinematics.}. 
 
 A combination of the best fitting SSP templates is determined and averaged to make a template model for each of the components and the fractional contribution of each model to the observed spectrum is optimized in order to construct the total fit. The top panel of Fig.~\ref{fig:how_works} shows that for bins dominated by lens light, the total fit is almost entirely constructed from SSP templates at redshift $z$\,=\,0.222. In contrast to this, bins such as the one shown in the bottom panel are composed of a relative contribution of the lens, source and gas templates.
 
 The addition of the two gas components to account for the Balmer emission lines and the oxygen doublet allows us to obtain a further  improvement to the fit with significantly reduced residuals. This can be seen in the second and third panels of Fig.~\ref{fig:zoomed_process} when compared with the first panel. The emission features that are not well fit by the 1 star model, such as the deep absorption wings on either side of the H-$\delta$ emission line (at $\sim$6600~\r{A} in the observed frame), are now present in the total fit of the 2 star + gas model.
 
\subsection{Kinematic Results}
\label{sec:kin results}

\begin{figure}
	\includegraphics[width=\columnwidth]{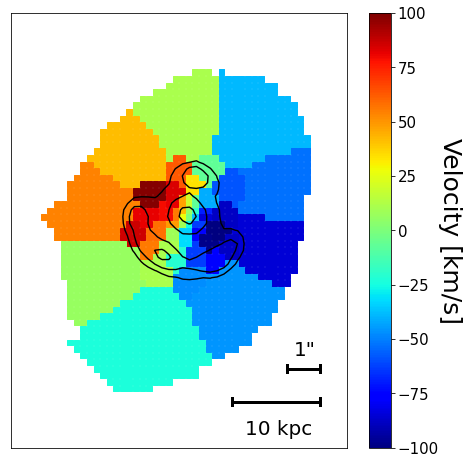}
    \includegraphics[width=\columnwidth]{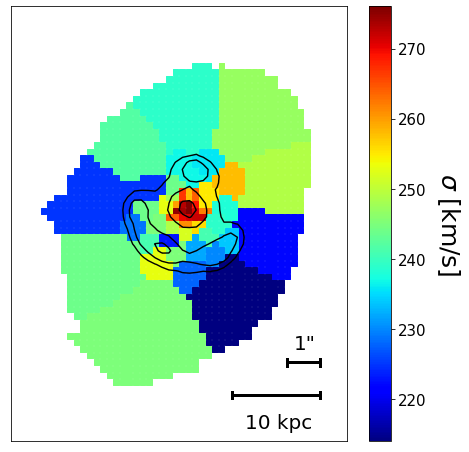}
    \caption{Velocity and velocity dispersion maps for an $\sim$\,8$\times$10\,arcsec$^2$
    region of the Jackpot, overlaid on contours of the galaxy flux. The upper panel shows a clear signature of rotation along the major axis and the lower panel shows a falling velocity dispersion gradient. }
    \label{fig:v_map}
\end{figure}

Fig.~\ref{fig:v_map} shows the derived velocity map for an $\sim$ 8$\times$10\,arcsec$^2$ area and clearly displays the axis of rotation of the lens galaxy, giving maximum velocities of $\sim \pm 100$\,km\,s$^{-1}$ about a kinematic axis with a misalignment with the photometric axis of the order 10\textdegree, dependent on the radius at which they are measured\footnote{The photometric position angle was measured using the {\sc MgeFit} package \citep[][]{Cappellari02}. The outer edge of the Voronoi binning is defined by a low-surface brightness envelope that is almost orthogonal to the true major axis of the galaxy.}.

This rotation has been hinted at in previous works, with \citet{Sonnenfeld12} noting evidence for some rotation in their analysis; however, the paper states that the stellar kinematics of the lens galaxy are dominated by pressure support, rather than rotation. Our data confirms that the kinematics are indeed dispersion dominated, with $v_{\rmn{rot}}^{2}\ll\sigma^{2}$. Fig.~\ref{fig:v_map} demonstrates the way in which our high-area-coverage MUSE data and multiple-component fitting allow us to fully map the 2D-kinematic properties of the Jackpot lens galaxy, out to a much greater radius than previous single-slit studies, allowing us to now fully characterise the rotation proposed by \citet{Sonnenfeld12}.

Fig.~\ref{fig:v_map} also shows the velocity dispersion map of the lens galaxy and illustrates the way in which $\sigma$ falls from $\sim280$\,km\,s$^{-1}$ at the centre of the lens to $\sim230$\,km\,s$^{-1}$ at a radius of $\sim$\,2\,arcsec, and highlights a discernible velocity dispersion gradient in the inner region. 

\begin{figure}
	\includegraphics[width=\columnwidth]{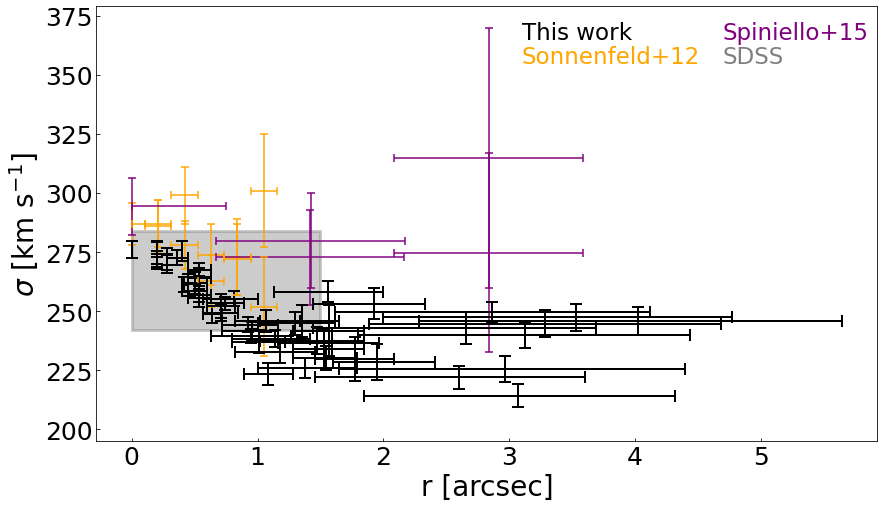}
    \caption{The velocity dispersion of each Voronoi bin as a function of radius from this work. For comparison, we show measurements from \citet{Spiniello15}, \citet{Sonnenfeld12}, and SDSS \citep[][]{auger09}. 
    Our data are broadly consistent with previous measurements in the inner regions, but show a clear decline in velocity dispersion towards a larger radius, where the precision of earlier datasets is lower. }
    \label{fig:kinematic_comparisons}
\end{figure}

Fig.~\ref{fig:kinematic_comparisons} shows the radial velocity dispersion profile derived from this study, as compared with that of \citet{Spiniello15}, \citet{Sonnenfeld12} and \citet{auger09}. Our measurements agree with those of previous studies in the regions with overlapping coverage (i.e. the inner $\sim$2\,arcsec). In the regions at larger radii than this, our measured profile exhibits the “quadrupole” structure seen in the velocity dispersion map in Fig.~\ref{fig:v_map}, with velocity dispersions that are lower on the major axis and higher on the minor. This behaviour is still found if we adopt very large bins in the outer regions; we further discuss the implications of this result in Section~\ref{sec:dynamicallysimple}.

\section{Dynamical Modelling with {\sc jam}}
\label{sec:modelling}

In this section, we model the spatially-binned kinematic measurements using the anisotropic Jeans model approach, as implemented in {\sc jam} \citep[][]{Cappellari08, cappellari12}. We impose a robustly constrained aperture lensing mass to reduce the freedom in the models. By optimising the model parameters as described in this section, we maximise the likelihood of the observed root-mean-square velocities, $v_{\text{rms}}$, obtained from the observed velocities and velocity dispersions in Section~\ref{sec:kin results} and given as $v_{\text{rms}}\,=\,\sqrt{v^2 + \sigma^2}$.

As seen by the extended envelope in Fig.~\ref{fig:data}, and further reflected in Fig.~\ref{fig:v_map}, we measure our stellar kinematics out to $\sim$~10\,kpc ($\sim$~2.7\,arcsec) from the centre of the lens. However, preliminary tests of our dynamical models showed indications that the Jackpot is not dynamically simple at greater radii, and we note here the possible evidence for tidal interactions found in \citet{Sonnenfeld12}. This presents a fundamental limitation of how well we can model the lens kinematics in the outer region, as this component cannot be accommodated by our simple dynamical models, which are limited to oblate axisymmetric mass and luminosity distributions. If indeed just tidal debris, we do not expect this to be very dominant in mass, but including the tracers in this region in the modelling could bias the recovery of the mass components of interest.

\begin{figure}
	\includegraphics[width=\columnwidth]{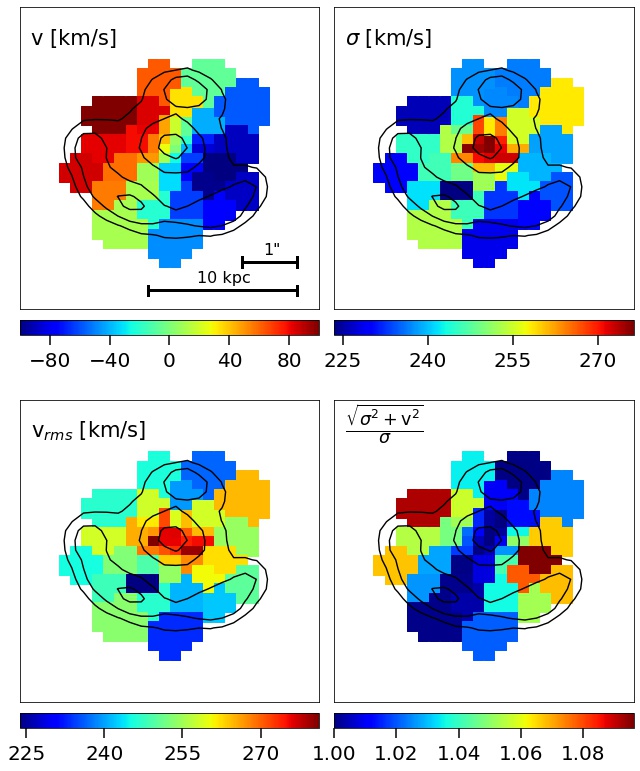}
    \caption{The measured kinematics from the restricted data, corresponding to the region used for dynamical modelling. The final panel shows the fractional increase of the second velocity moment caused by ordered rotation.}
    \label{fig:kin_maps_trunc}
\end{figure}

As a result, we determine that our Jeans models are not appropriate to model the lens kinematics at greater radii, and as such, all dynamical models described herein are fit to the exclusion of our nine outermost bins. We thus opt to restrict the radius for dynamical model predictions to 1.95\,arcsec, which is still sufficient to allow projected mass slope measurements at the Einstein radius. We further discuss this choice and its implications in Section \ref{sec:dynamicallysimple}. The measured kinematics for this restricted region are shown in Fig.~\ref{fig:kin_maps_trunc}, the final panel of which shows the fractional increase of the second velocity moment by ordered rotation, which further demonstrates that the galaxy is dispersion dominated within the fitted radius.

\subsection{Mass Model}
\label{sec:mass_model}

To construct our dynamical mass models, the mass distribution of the lens galaxy is represented as a combination of the stellar mass de-projected from the observed light (with constant stellar mass-to-light ratio), a spherical dark matter halo, and an excess central mass component, as further detailed in this section. The model components are parameterised in terms of their fractional contribution to the lensing mass inside the Einstein radius. Therefore, all models explored are consistent with the lensing configuration.

\subsubsection{Stellar Mass}
\label{sec:stellar_mass}

 \begin{figure*}
	\includegraphics[width=2\columnwidth]{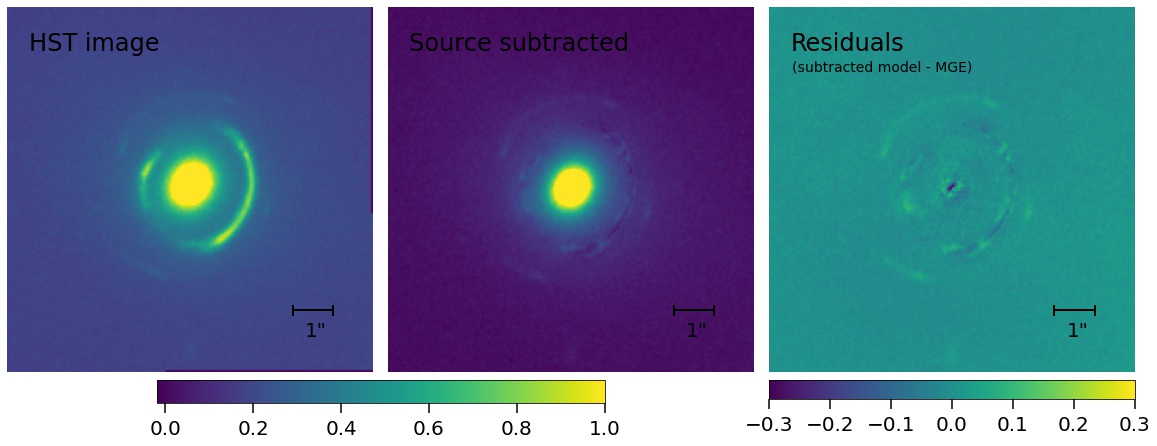}
    \caption{The F814W-band HST image, the HST image with the parameterised source model subtracted and the residuals from the MGE model subtraction. Each panel corresponds to a field of view of 9\,arcsec on a side.}
    \label{fig:mge_plots}
\end{figure*}

We follow the common approach of de-projecting the luminosity density using a multi-gaussian expansion (MGE) \citep[as per][]{Emsellem94, Cappellari02} fit to high-resolution imaging from HST, where the MGE projected surface brightness is given as

\begin{equation}
\label{eq:MGE}
\Sigma(R',\theta') = \sum_{j=1}^{N}\frac{L_{j}}{2\pi\sigma_{j}^{2}q'_{j}}~ \mathrm{exp}\left [  -\frac{1}{2\sigma_{j}^{2}} \left ( {x'_{j}}^{2} + \frac{{y'_{j}}^{2}}{{q'_{j}}^{2}}\right )\right ],
\end{equation}

where $(R', \theta')$ are the polar coordinates on the plane of the sky $(x', y')$. $N$ is the number of Gaussian components with total luminosity given by $L_j$, observed axial ratio of $0 \leq q'_j \leq 1$ and width along the major axis of $\sigma_j$, as per \citet{Cappellari02}.

As this method would be unreliable in the presence of the bright Einstein arc, we fit our MGE to an image from which the arc has been subtracted using a lens reconstruction model,
as is described in Section~4 of \citet{etherington22}. Specifically, we use an image from HST (F814W-band) from which a {\sc PyAutoLens}-fitted \citep[][]{nightingale18, nightingale21} parameterised source model has been subtracted. 
This image, with a rest-frame wavelength of $\sim$6700\,\AA, is not expected to be sensitive to any modest variations in age and metalicity, and therefore its luminosity in this band is assumed to trace the stellar mass surface density reasonably well (in shape, but not normalisation) in the absence of any IMF gradients. The parameterised model assumes the source galaxy to be well-described intrinsically by a smooth S\'ersic-profile galaxy. Although this method yields a less precise source-subtraction than the alternative pixelised-source approach, it is preferred here to avoid overfitting of the lens galaxy light.

As implemented in the {\sc MgeFit} code, the luminosity profile is measured in a number of elliptically defined sectors of this residual image of Jackpot and the MGE fits a series of Gaussians to the profile, describing the intensity and shape of the total surface brightness. The PSF of the HST image was approximated with a single Gaussian of 0.1\,arcsec FWHM; however, little difference was found in the recovered parameters when this value was varied within reasonable limits. The projected mass density in stars is then assumed to be proportional to the projected luminosity density, i.e. with a constant stellar mass-to-light ratio. Table \ref{tab:table1} presents the stellar mass density MGE model for the best fitting model parameterisation of the Jackpot lens, as described in Section~\ref{sec:results}. 

\begin{table}
	\centering
	\caption{MGE$_{\star}$ for the Jackpot galaxy. The columns represent, left to right, the projected surface mass density multiplied by the best fitting stellar fraction parameter in the free $\gamma$ {\sc jam} models, the MGE width, and axis ratio.}
	\begin{tabular}{ccc}
		\hline
		$\Sigma_{\star}$ $\times$ f$_{\star}$ & $\sigma$ & q\\
        $[\rmn{M}_{\odot}$/pc$^{2}]$ & [arcsec] &  \\
		\hline
		7609.12 & 0.075 & 0.984\\
		6370.03  & 0.197 & 0.896\\
		2715.53 & 0.442 & 0.821\\
		\z751.32 & 0.890 & 1.000\\
		\z268.85 & 2.366 & 1.000\\
		\hline

	\end{tabular}

    \label{tab:table1}
\end{table}

Fig.~\ref{fig:mge_plots} shows the F814W-band HST image, the HST image with the smooth S\'ersic-profile source model subtracted, and the residuals obtained from subtracting the MGE luminosity profile from the source-subtracted model. In the absence of the source, the lens is parameterised simply by an ellipse with a S\'ersic profile. We note that whilst the source-subtracted model fails to reproduce all of the observed features in the arcs that reflect real structures in the source, this treatment is sufficient to allow the natural robustness of {\sc MgeFit} to follow the true luminosity distribution of the lens galaxy. 

\begin{figure}
	\includegraphics[width=0.9\columnwidth]{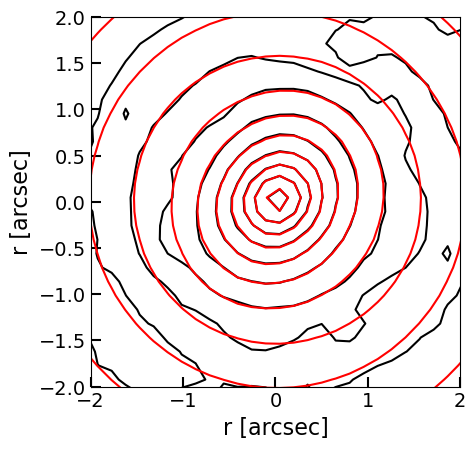}
    \includegraphics[width=0.9\columnwidth]{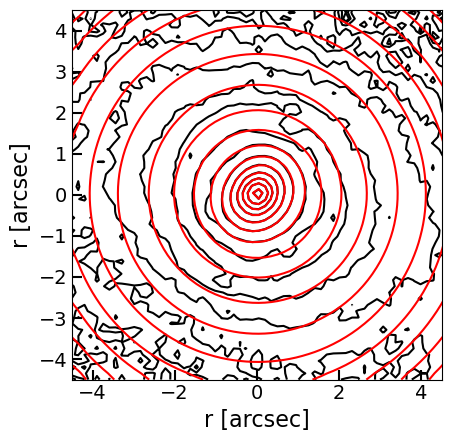}
    \caption{Wide-field and zoom-in isophote plots of the source-subtracted image of Jackpot (black). The contours of the MGE model surface brightness are overlaid in red. The MGE is a good fit in the central regions, but deviations are evident at large radii due to the outer stellar envelope. 
    }
    \label{fig:mge_contours}
\end{figure}

Fig.~\ref{fig:mge_contours} shows an isophote plot of both the source-subtracted image of Jackpot and the MGE model. The MGE model surface brightness is in good agreement with that of the galaxy in the inner $\sim$ 2 arcsec (i.e. where our kinematics are measured), but poorly reproduces the shape of the observed isophotes beyond this, where the light from the diffuse outer envelope becomes significant, as also seen in Fig.~\ref{fig:data}. We note the presence of this same effect in Fig.~1 of \citet{posacki15}, but find that in comparison, our treatment of the HST image works in reducing the contamination from the bright arcs and improves the model fit in the inner regions.

\subsubsection{Dark Matter Halo}
\label{sec:DM_halo}

A similar process to the one described in Section~\ref{sec:stellar_mass} is followed to obtain a second series of Gaussians describing the galaxy DM surface density. Here, the DM halo is assumed to be spherical and have density well-described by a generalized NFW profile \citep[gNFW;][]{zha096} of the form

\begin{equation}
\label{eq:2}
\rho_{\text{DM}} = \frac{1}{(r/r_{\text{s}})^{\gamma} (1 + r/r_{\text{s}})^{3-\gamma}},
\end{equation}

where $r$ is the physical radius and with $\gamma$\,=\,1 corresponding to the original NFW slope. Here $r_{\rm s}$ is the scale radius, fixed at 100\,arcsec based on reasonable assumptions for the virial radius of the order 600\,kpc \citep[as per][]{Gavazzi07}, a halo concentration parameter of c$_{\rmn{vir}}\approx6$ \citep[from a mass-concentration relationship from][]{maccio08} and given the NFW halo density profile. It was found that varying this value slightly (i.e. by 20\%) had very little effect on the recovered kinematic estimates, as is expected since the observational constraints are well inside $r_{\rm s}$.

In our use of the gNFW profile, we are not assuming any specific physical origin of any difference with respect to the NFW profile, but a slope of $\gamma>1$ could represent, for example, a contraction in response to the baryonic mass \citep[e.g][]{blumenthal86, gnedin04}.

\subsubsection{Excess Central Mass}
\label{sec:BH}

We include an additional `excess' central mass component, $m_{\text{cen}}$, through the mechanism that {\sc jam} uses to model a central black hole. In our implementation, this is understood to subsume any real point mass, i.e. a black hole, as well as any centrally concentrated mass in excess of a constant stellar mass-to-light ratio. This excess mass is modelled as an additional, very small Gaussian component and is explored in the range $0 \leq$ $m_{\text{cen}}$ $\leq 7\times10^{10}$ M$_{\odot}$, giving the mass model a further degree of flexibility at smaller radii. We further discuss the simplification of a mass component of $\sim$zero radius in Section~\ref{sec:BH?}.

\subsubsection{Model Normalisation}
\label{sec:normalisation}

In order to derive the total-mass surface density, the Gaussians resulting from the MGE fits to the luminous and dark components, as described in Sections~\ref{sec:stellar_mass} and \ref{sec:DM_halo} respectively, are combined with the excess central mass described in Section~\ref{sec:BH}. A projected luminous fraction (i.e. the stellar mass as a fraction of the Einstein mass from lensing) is explored in the range $0 \leq f \leq 1$. We reduce the freedom in the dynamical models by rigidly enforcing a robustly-constrained lensing mass $m_{\rm E}$ at the Einstein radius, normalising the MGE such that

\begin{equation}
\label{eq:3}
\frac{m_{\star}(<\theta_{\text{E}})}{m_{\text{E}}} + \frac{m_{\text{DM}}(<\theta_{\text{E}})}{m_{\text{E}}} + \frac{m_{\text{cen}}}{m_{\text{E}}} = 1\ ,
\end{equation}
where 
 $m_{\star}(<\theta_{\text{E}})$, $m_{DM}(<\theta_{\text{E}})$ and $m_{\text{cen}}$ are the projected mass contributions from stars, dark matter and the central excess, to the total Einstein-aperture lensing mass. Given the redshifts of the source and lens in the Jackpot system, the measured $\theta_{\text{E}}$\,=\,1.397\,arcsec \citep[][]{Collett14}  yields $M_{\rm E}$\,=\,3.08\,$\times$\,10$^{11}$\,$M_{\rm \odot}$.

\subsection{Anisotropic Modelling and Parameter Search}
\label{sec:method}
The main goal of this study is to measure the slope of the total 2D-projected mass profile. To minimise any bias in this quantity, we adopt a model that is sufficiently flexible to reproduce the observed kinematics.

The normalised MGE descriptions of the surface brightness and the total-mass surface density, along with the variable parameters described in this section, are used to calculate a prediction of the projected $v_{\text{rms}}$ field for an anisotropic axisymmetric galaxy model. The prediction includes smoothing by the MUSE PSF approximated as a single Gaussian with FWHM of 0.5\,arcsec\footnote{The smoothing imposed by the Voronoi binning does not exceed the PSF until a mean radius of $\sim1.42\,$arcsec, beyond which point the $v_{\text{rms}}$ field is slowly varying on the relevant scales.}. The predicted second moments are calculated at the luminosity-weighted Voronoi bin centres for comparison with the observed data.

We explore parameter space for two distinct model sets with velocity dispersion ellipsoids aligned with the cylindrical ($R,z$) polar coordinate system \citep[the cylindrical {\sc jam} method,][]{Cappellari08, cappellari12}; a set of models with the DM density slope as a free parameter, and a further set of models with the DM density fixed as a NFW profile. Within the cylindrically-aligned paradigm, the orbital anisotropy parameter is defined as $\beta_z = 1 - \overline{v_{z}^{2}}/ \overline{v_{R}^{2}}$ \citep[][]{Cappellari08}. 

We explore the parameter space using a Markov chain Monte Carlo (MCMC) Ensemble sampler, as described by \citet{emcee}, and summarise the free parameters of our models described above as follows:

\begin{itemize}
  \item $\gamma$, the DM density power law slope. Values in the range $0.5 \leq \gamma \leq 3$ are explored, with $\gamma\,=\,1$ describing the standard NFW slope;
  \item $\beta$, the orbital anisotropy parameter which, in the cylindrically-aligned case, describes the ratio of the radial velocity dispersion to the vertical component. Here we consider values of $-0.6 \leq \beta \leq 0.6$, where a negative value of beta indicates a relatively larger vertical velocity dispersion.
  \item $f_{\star}$, the stellar mass as a fraction of the total lensing mass, explored from $0 \leq$ $f_{\star}$ $\leq 1$;
  \item $m_{\text{cen}}$, the excess central mass. Here, the mass range $0 \leq$ $m_{\text{cen}}$ $\leq 7\times10^{10}$ M$_{\odot}$ is explored.
  \item $i$, the galaxy inclination, with a lower limit of 35\textdegree~imposed by the minimum observed axial ratio of the MGE Gaussians describing the distribution of the kinematic-tracer population. This is defined such that an inclination of 90\textdegree~corresponds to the edge-on case. 

\end{itemize}

We define the prior probability density function (PDF) for all of our parameters with flat priors in $\gamma$, $\beta$, $f_{\star}$, $m_{\text{cen}}$ and $i$, and impose physically motivated constraints on the extreme values as described above. 

\subsection{Results}
\label{sec:results}

\begin{figure*}
	\includegraphics[width=1.8\columnwidth]{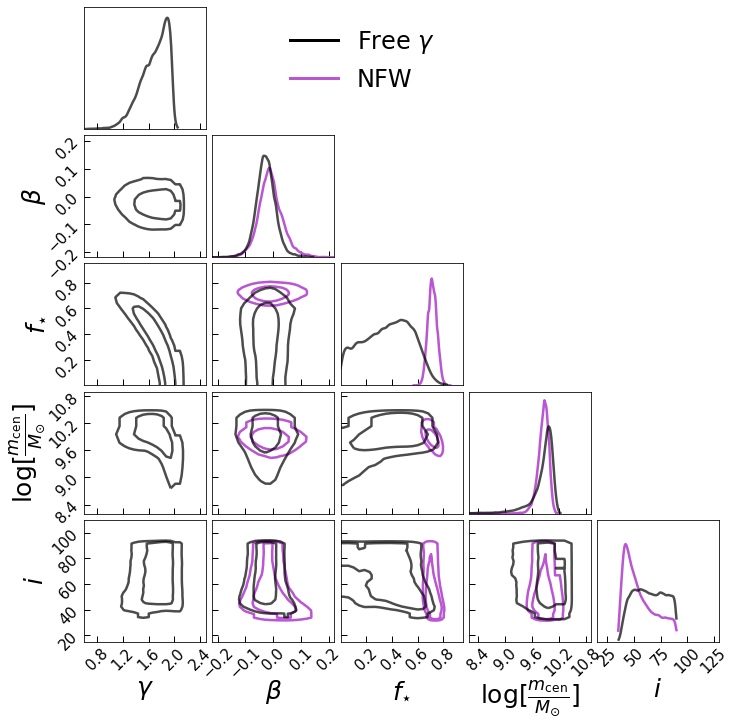}
    \caption{The posterior PDF for the free $\gamma$ and NFW model parameters. The contours show the 68 and $95\%$ confidence regions. The parameters explored are: the inner slope of the DM density profile, $\gamma$; the orbital anisotropy parameter, $\beta$; the stellar mass as a fraction of the total lensing mass, $f_{\star}$; any central mass in excess of a constant stellar mass-to-light ratio, $m_{\text{cen}}$; the galaxy inclination, $i$. The diagonal plots show the marginalised posterior densities for each parameter.}
    \label{fig:big_jam_contours}
\end{figure*}

Using the likelihood derived from the predicted and observed $v_{\text{rms}}$, and given the priors above, we sample the posterior PDF for the five model parameters. Fig.~\ref{fig:big_jam_contours} shows the marginalised parameter constraints, which are given in Table~\ref{tab:table2}. The results of our modelling can be summarised as follows:

\begin{table}
	\centering
	\caption{The median and 68\% confidence bounds for the model parameters from both of our model sets.}
    \def\arraystretch{1.75}%
	\begin{tabular}{cccccc}
		\hline
		Model & $\gamma$ & $\beta$ & $f_{\star}$ & $m_{\text{cen}}$[$10^{9}$M$_{\odot}$] & $i$[\textdegree]\\
		\hline
		Free $\gamma$ & $1.73\substack{+0.17 \\ -0.26}$\ & $-0.03\substack{+0.03 \\ -0.03}$ & 
        $0.38\substack{+0.19 \\ -0.23}$ & $8.23\substack{+2.56 \\ -3.67}$ & 
        $64\substack{+17 \\ -16}$\\
		NFW  & ($=1$) & $-0.01\substack{+0.04 \\ -0.04}$\ & 
        $0.71\substack{+0.04 \\ -0.03}$\ & $7.31\substack{+1.82 \\ -1.82}$\ &
        $51\substack{+21 \\ -10}$\\

		\hline 
	\end{tabular}
    \label{tab:table2}
\end{table}

\begin{figure}
	\includegraphics[width=\columnwidth]{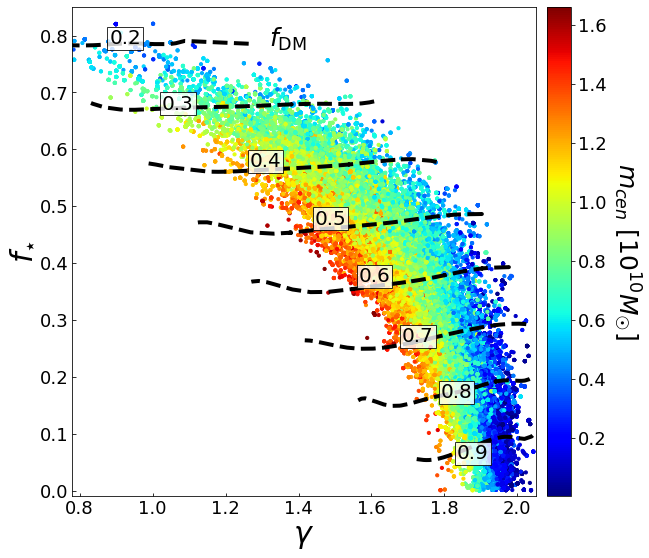}
    \caption{ The stellar mass as a fraction of the total lensing mass for all free $\gamma$ models, as a function of the DM density slope. Coloured points represent the preferred excess central mass and the black dashed line denotes contours for the corresponding DM mass fraction.}
    \label{fig:f_vs_gamma_mbh}
\end{figure}

\begin{itemize}
    \item The preferred excess central mass is well-constrained to be $\sim8\times10^{9}$\,M$_{\odot}$ for the free $\gamma$ models, and $\sim7\times10^{9}$\,M$_{\odot}$ for the NFW models. If this component indeed represents only a central black hole, this would make it somewhat over-massive given the galaxy properties. For a galaxy such as the Jackpot lens with a central velocity dispersion of $\sim280$\,km\,s$^{-1}$, from the standard black hole mass vs. sigma relation \citep[][]{VanDenBosch16}, one would expect a central black hole with mass $\sim1.6\times10^{9}$\,M$_{\odot}$ (with a scatter of $0.49\pm0.03$). Given this discrepancy, it seems unlikely that all excess central mass is contributed by a black hole. We also note the potential for a systematic overestimation of the central mass component, as described in Appendix~\ref{sec:appendix} and further discuss the implications of this in Section~\ref{sec:BH?}.
    
    \item Our inferred DM density slope, taken from the model set with $\gamma$ as a free parameter, is $1.73\substack{+0.17 \\ -0.26}$, significantly steeper than an unmodified NFW profile (i.e. $\gamma\,=\,1$). Such a slope could represent a baryon-contracted halo appropriate to a massive galaxy \cite[][]{sonnenfeld21}. We find that models with a slope flatter than NFW are strongly disfavoured, but not disallowed. 

    \item In the free $\gamma$ models, there is a strong degeneracy between $f_{\star}$ and $\gamma$, which prevents us from obtaining closed inference on the stellar fraction. Instead, a range of $0.15<f_{\star}<0.57$ is somewhat weakly favoured. Thus, the inferred mass budget inside the Einstein radius for the models with a free DM density slope is $m_{\star}:m_{\text{DM}}:m_{\text{cen}}\,=\,0.38:0.59:0.03$, albeit with substantial uncertainty. For the models with a fixed NFW-like DM density, this parameter is much more tightly constrained and the mass budget is $0.71:0.27:0.02$. 
    The expected stellar mass fraction under the assumption of a Chabrier IMF is $f_{\star}^{\text{Chab}}\,=\,0.26\pm0.07$, whilst with a Salpeter IMF we expect $f_{\star}^{\text{Salp}}\,=\,0.46\pm0.13$ \citep[][]{auger09}. Our preferred stellar mass fraction for the free $\gamma$ models is broadly consistent with that of a Salpeter IMF. For the NFW models, we find a stellar mass fraction heavier than the predictions of a Salpeter IMF and that is inconsistent with a Chabrier IMF.

    \item In the free $\gamma$ case, a DM mass fraction of $f_{\text{DM}}\,=\,0.59\substack{+0.24 \\ -0.19}$ was inferred. This sits in good agreement with that of \citet{Gavazzi08}, who found a surprisingly high DM mass fraction inside the effective radius (2.0\,arcsec for the Jackpot) of $f_{\text{DM}}(<R_{\text{eff}})\,=\,0.73\pm0.09$.

    \item The velocity ellipsoid is strongly constrained to be nearly isotropic in both model sets, with orbital anisotropy parameters of $\beta\,=\,-0.03\pm0.03$ and $\beta\,=\,-0.01\pm0.04$ for the free $\gamma$ and NFW cases respectively, which is consistent with the low values typically found, e.g. for $\sigma > 200$\,km\,s$^{-1}$ galaxies in the ATLAS$^{\text{3D}}$ survey \citep[][]{Cappellari11}. The model likelihood is rather insensitive to the orbital anisotropy, although this is perhaps to be expected given the context of the assumed cylindrically-aligned coordinate system.
    
    \item The stellar mass as a fraction of the total lensing mass is highly degenerate with the free DM density power law slope; this is to be expected as, to first order, either reducing the stellar fraction at the expense of the DM fraction, or flattening the DM profile, act to effectively predict larger root-mean-square velocities at larger radius. Despite $\gamma$ being free to explore values up to a slope of 3, we see that our models do not exceed a slope of $\sim2$ due to the hard prior imposed on the lower limit of the stellar fraction, and the strong anti-correlation between these two parameters. This behaviour is further demonstrated in Fig.~\ref{fig:f_vs_gamma_mbh}, which shows that the models with the steepest DM profile are also the models with the lowest stellar mass and smallest excess central mass. The apparent cut-off at $\gamma\sim2$ corresponds to the case where the DM component accounts for the totality of the dynamical mass; there is simply no further flexibility in the mass budget for a steeper $\gamma$ to be explored.
    
    \item In the model set with a fixed NFW-like DM density slope, the stellar fraction is anti-correlated with $m_{\text{cen}}$. We see that the models with a high stellar fraction prefer smaller excess central masses as, in the absence of flexibility from a free DM slope parameter, both components act to account for any compact additional central mass, so are free to compensate one another. The model set with a free $\gamma$ does not demonstrate this behaviour, as there is instead the freedom for interplay between the three mass components in the models, constrained by the total lensing mass. In these cases, the greater concentration of DM in the central regions reduces the necessity for such a large excess central mass component, and suppresses the stellar contribution.   

    \item We found the galaxy inclination to be completely unconstrained in the free $\gamma$ models, and only somewhat constrained in the NFW-like models, with an inferred inclination value of $51\substack{+21 \\ -10}$. More `edge-on' inclinations up to 90\textdegree~were not strongly excluded, but were instead disfavoured. In both cases, the inclination shows no significant covariance with the parameters of interest.
 
\end{itemize}

\begin{figure*}
	\includegraphics[width=2.08\columnwidth]{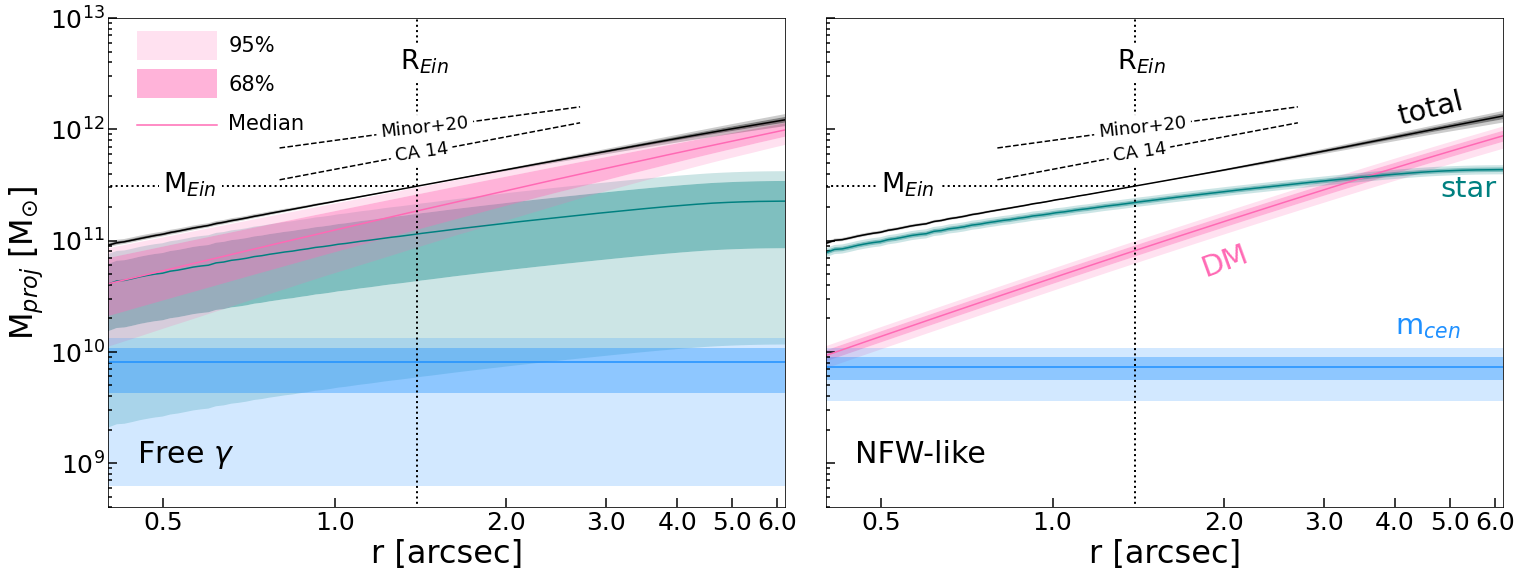}
    \caption{The 2D-projected mass profiles for each individual mass component: the stellar (green), DM (pink) and excess central mass (blue) components, as well as the total projected mass profile (black), for both the model with a free DM slope parameter (left) and the model with NFW-like DM (right). To enable comparisons, the slopes derived by \citet{Collett14} and \citet{Minor21} have also been plotted. The vertical and horizontal dotted lines denote the Einstein radius and mass, respectively.}
    \label{fig:proj_mass}
\end{figure*}

Fig.~\ref{fig:proj_mass} shows the 2D-projected mass profiles for each of the individual mass components described in Section \ref{sec:mass_model}, for both the free $\gamma$ and NFW cases.

The large range of profiles and relative contributions in each model demonstrate the way in which the free $\gamma$ models have a large degree of flexibility to effectively `trade off' mass components at different characteristic radii. These models prefer a steep ($\gamma>1$) DM density slope, and hence their range of DM projected mass profiles here are shallower than seen in the NFW models. The median contribution of the DM mass component is significantly greater for $r\lesssim3$\,arcsec than in the NFW models, as this model set has the freedom to allow for a greater concentration of DM mass in the central regions whilst still being constrained by the total lensing mass, thus dominating the total mass in the centre of the galaxy.

Conversely, the models with a DM density slope fixed at NFW-like show DM mass contributions that do not dominate the total projected mass until well outside the Einstein radius ($r\gtrsim4$\,arcsec), and instead prefer a more substantial stellar mass fraction. Indeed, Fig.~\ref{fig:proj_mass} shows that, in the absence of a steep enough DM halo, the NFW-like models are almost solely dominated by the stellar contribution in the central regions. Notwithstanding uncertainty in the relative component contributions, the projected mass profile at the Einstein radius is tightly constrained by the data, and the flexibility afforded to the model parameters results in almost indistinguishable recovered total mass slopes. 

\begin{figure}
	\includegraphics[width=\columnwidth]{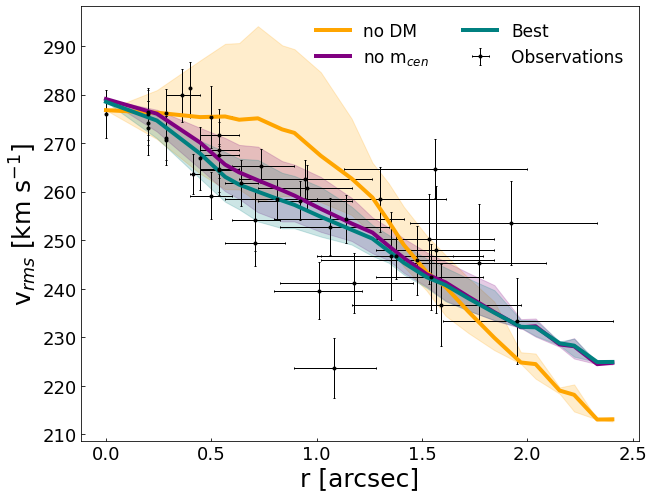}
    \caption{The mean and azimuthal range of the root-mean-square velocity predictions from our best model (green) compared to the best possible model without an excess central mass (purple), and the best model predictions without accounting for the mass in DM (orange). Also shown are the measured $v_{\text{rms}}$ from each Voronoi bin as a function of radius (black). The horizontal error bars of the observed data represent the width of the Voronoi bins. The model set without a DM halo fails to predict the observed kinematics, while the model set without an excess central mass can successfully reproduce the kinematics, although in this case the recovered stellar mass and DM halo profile are unreasonable (see text).}
    \label{fig:best_we_can_do}
\end{figure}

\begin{figure*}
	\includegraphics[width=1.8\columnwidth]{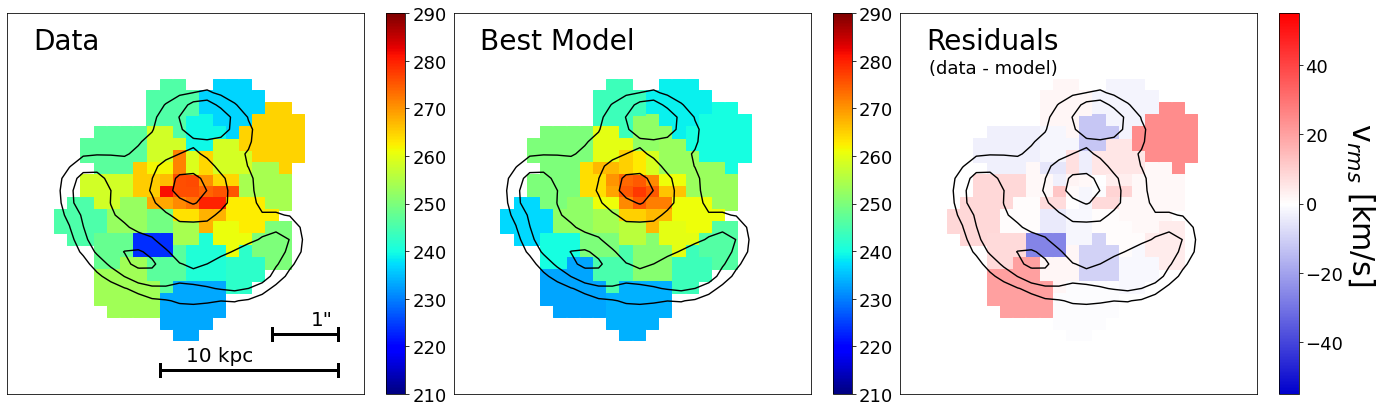}
    \includegraphics[width=1.8\columnwidth]{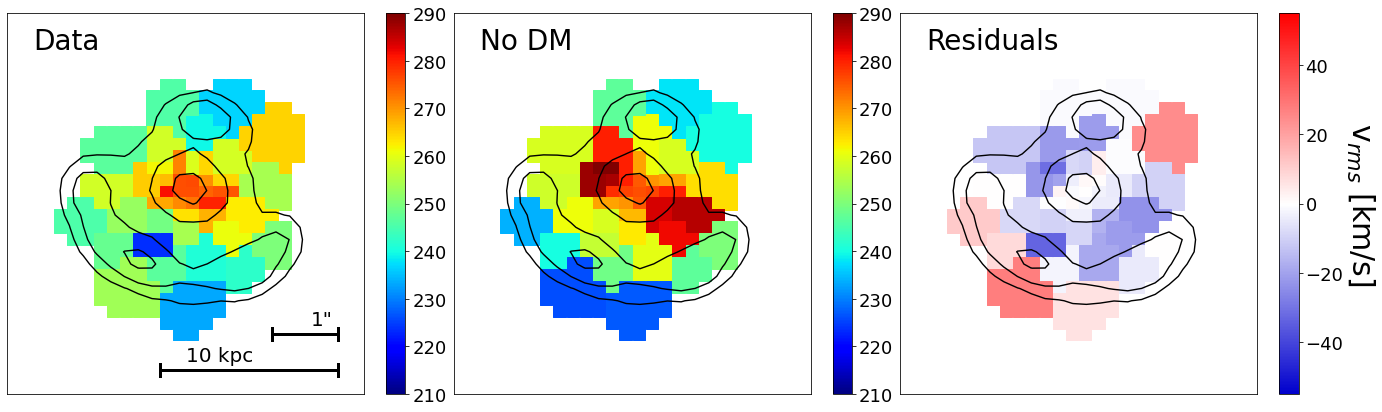}
    \includegraphics[width=1.8\columnwidth]{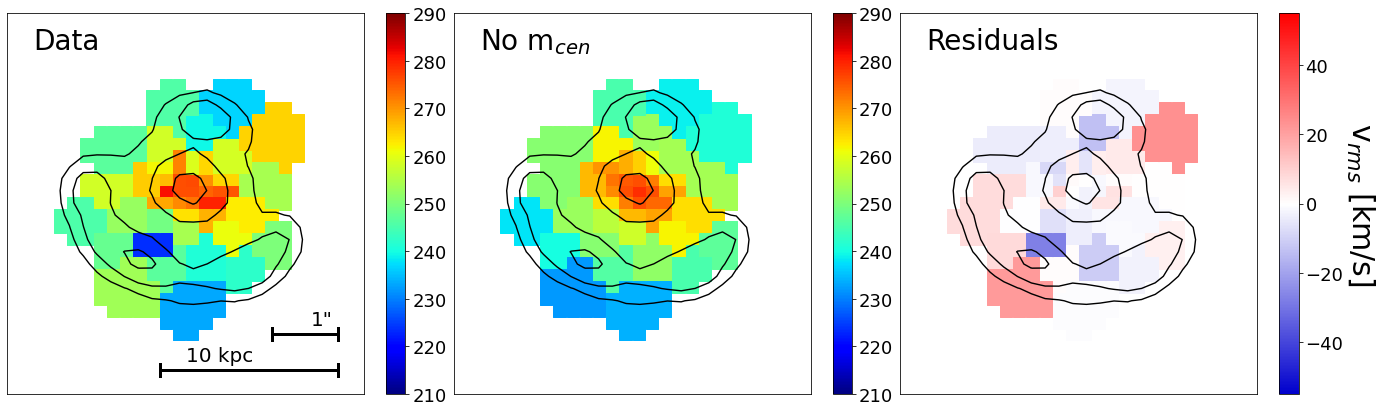}
    \caption{Kinematic maps for the root-mean-square velocities, $v_{\text{rms}}$, obtained from the data, alongside the best fitting models for each scenario described in Fig. \ref{fig:best_we_can_do}. Also shown are the model residuals. The best fitting model exhibits a clear $v_{\text{rms}}$ gradient and the residuals are $\sim \pm 40$\,km\,s$^{-1}$. The no-DM model exhibits higher residuals than the best fitting model and completely fails to predict the observed root-mean-square velocities obtained from the observed data.}
    \label{fig:model_maps}
\end{figure*}

Figs.~\ref{fig:best_we_can_do} and \ref{fig:model_maps} show the $v_{\text{rms}}$ predictions from our best model, which reproduces the data well over the full range of measurements, compared to predictions of a model without a DM halo, that attributes the gravitational potential solely to luminous matter and any additional central mass, and of a further model with no excess central mass component. 
As we only constrain our total mass around the Einstein radius, our three contributing mass parameters (stellar, DM and central) are free to compensate one another, e.g. in the absence of the excess central mass required to reproduce the observed higher velocities in the innermost regions (where the $m_{\text{cen}}$ component would dominate), consequent adjustment between the DM and stellar components are necessary to produce the steep inner $v_{\text{rms}}$ profile.
The kinematic predictions from the best fitting dynamical model exhibit a clear gradient in $v_{\text{rms}}$, with values ranging from $\sim280$\,km\,s$^{-1}$ at the centre of the lens to $\sim230$\,km\,s$^{-1}$ in the outer regions, in close agreement with the observed kinematics, with relatively small residuals. The models that do not include all three mass components, however, unsurprisingly fail to predict the observed kinematics whilst simultaneously constraining model parameters that are realistic, as the models are not sufficiently flexible. Fig.~\ref{fig:model_maps} shows that the angular structure of the model without an excess central mass component is similar to that of the best model, while the no-DM model demonstrates prominent high $v_{\text{rms}}$ lobes along the major axis. The model without a DM halo substantially overestimates the $v_{\text{rms}}$ in a range 0.2\,arcsec~$\lesssim r\lesssim$~1.2\,arcsec. This is a result of the total mass distribution of the lens being solely constrained by the centrally-concentrated luminous MGE in this case, and lacking an extended mass component. 
The model without an excess central mass component successfully reproduces the observed kinematics at all radii, but does so at the expense of requiring an unrealistic stellar mass fraction ($f_\star\,=\,0.09$) and a very steep DM density profile ($\gamma\,=\,1.97$). This is consistent with the $\gamma$-$f_\star$ panel in Fig.~\ref{fig:big_jam_contours} and the dark blue data points in Fig.~\ref{fig:f_vs_gamma_mbh}. While formally consistent with the lensing and dynamics, relative to the tabulated values of \citet{auger09}, a model with such a low stellar mass fraction would imply a stellar IMF that is a factor of 2-3 lighter even than the Chabrier IMF.

\subsection{Total Mass Profile Slope}
\label{sec:profile_slope}

To compare our result to previous (non-dynamical) lensing studies of the Jackpot, we must relate our composite profile results to the power-law total density profile used in the lensing literature, \citep[e.g.][]{Collett14, Minor21, etherington22}\footnote{In practice, our composite profile is close enough to a power law that the two slopes are expected to be very similar.}. The sum of a stellar component and a dark matter halo does not yield a power-law total density profile, yet making this approximation has historically been acceptable, due to the so-called bulge-halo conspiracy \citep[][]{dutton14}: for lensing ellipticals, the sum of stellar and dark matter components is approximately isothermal over the length-scales probed by strong-lensing constraints.

To define an equivalent power law index  from our composite model, we use all of the Gaussian components for the profile (constructed as described in Section~\ref{sec:mass_model}) to compute the 2D-projected mass profile. For our free $\gamma$ and NFW model sets, the {\it local} slope of this profile, at the Einstein radius, is $1.03\pm0.03$ and $1.07\pm0.04$ respectively. As this is a profile of integrated mass, a value larger than unity corresponds to a shallower-than-isothermal density profile.

This quantity is not directly comparable to the lensing literature, as lensing does not strictly measure the local density slope at the Einstein ring. We instead calculate the power-law profile with the equivalent lensing effect as our inferred composite model.  \citet{collett_thesis} and \citet{kochanek20} showed that the slope inferred from lens modelling is sensitive to the radial derivative of the deflection angles at the location of the lensed images. The parameter we need to calculate is the dimensionless quantity $\xi$, which is well-constrained by lensing data as per \citet{kochanek20} and written as
\begin{equation}
\label{eq:7}
\xi = \frac{R_{E}\alpha^{\prime\prime}}{1 - \kappa_{E}},
\end{equation}
where $\kappa_{E}$ is the mean convergence at the Einstein radius and $\alpha^{\prime\prime}$ is the second derivative of the deflection profile at $R_{E}$. For models that use a power law relation with surface mass density $\propto r^{-\eta}$, this quantity is given by
\begin{equation}
\label{eq:8}
\xi = 2(\eta - 1).
\end{equation}

\begin{figure}
	\includegraphics[width=\columnwidth]{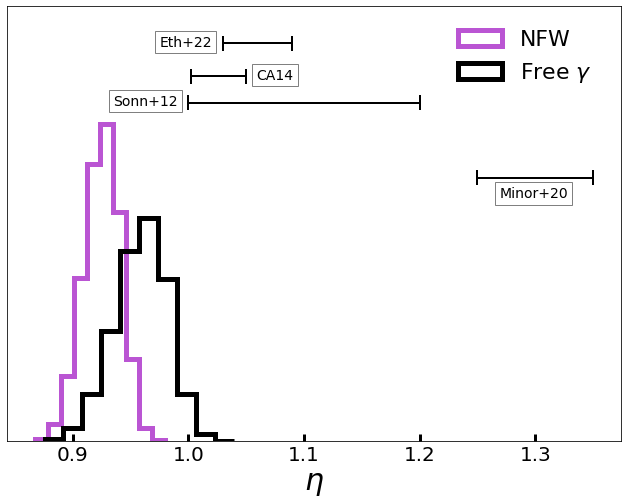}
    \caption{The 2D-projected total logarithmic density slope obtained from each of our two model sets; the free $\gamma$ model and the NFW-like model. The slopes derived by \citet{Sonnenfeld12}, \citet{Collett14}, \citet{Minor21} and \citet{etherington22} and their respective errors have been included for comparison.}
    \label{fig:hist_alpha}
\end{figure}

We therefore use our composite mass model to calculate $\xi$ from Equations~\ref{eq:7}, and convert this to a lensing-equivalent power-law slope $\eta$ using Equation~\ref{eq:8}. This gives us the value of $\eta$ that should be  measured from a lensing-only study for a mass profile with our best-fitting parameters. As shown in Fig. \ref{fig:hist_alpha}, sampling from the posterior distribution, we find a projected logarithmic density slope of $\eta\,=\,0.96\pm0.02$ for our free $\gamma$ models and $\eta\,=\,0.93\pm0.02$ for our NFW models. This indicates a density profile that is marginally shallower than the isothermal case, where $\eta$\,=\,1.

\section{Discussion}
\label{sec:discussion}

\subsection{Comparisons with the literature}
\label{sec:slope_comparisons}

The near isothermal density slope that we have measured is reasonable given previous studies of populations of massive elliptical galaxies. \citet{koopmans06, Grillo08, duffy10} all found that for the ensemble of lens galaxies, the average total density slope is approximately isothermal.

For the Jackpot lens specifically, our lensing-equivalent total density-profile slope of $\eta$\,=\,0.96$\pm$0.02 (for the free $\gamma$ model) is in good agreement with previous measurements of $1.00\pm0.03$, $1.1\pm0.1$, $1.03\pm0.02$, $1.06\pm0.03$ and $1.01\pm0.18$ from  \citet{Gavazzi08}, \citet{Sonnenfeld12}, \citet{Collett14}, \citet{etherington22} (lensing only) and \citet{etherington23} (lensing + dynamics), respectively \footnote{We do, however, note here that the \citet{Gavazzi08} and \citet{Sonnenfeld12} slopes are derived from a measure of the average slope between the two Einstein rings, not at the location of the first ring as is done in this paper, and the \citet{Collett14} constraint also includes the power-law slope's effect on the light profile of the second ring.}. Our model also does an excellent job of predicting the lensing deflection at the location of the second ring, with a deflection angle of $1.89 \pm 0.03$ arcseconds. This is $\sim$0.2 arcseconds less than the Einstein radius of the second ring, but a slight underestimate is to be expected since our model neglects the presence of mass in the first source. \citet{Collett14} inferred an SIS Einstein radius of $0.16\pm0.02$\,arcsec for the first source. Adding this to our deflection angle yields a second Einstein radius of $2.05 \pm 0.04$\,arcsec, entirely consistent with the $2.07 \pm 0.02$\,arcsec measured by \citet{Gavazzi08}. Additionally, we infer a DM density slope of $1.73\substack{+0.17 \\ -0.26}$ which is significantly steeper than an unmodified NFW profile (i.e. $\gamma\,=\,1$), but sits in good agreement with the lensing + dynamics DM profile slope of $1.7\pm0.2$ found for this galaxy by \citet{Sonnenfeld12}. 

Conversely, our derived total density slope is inconsistent  with that from the \citet{Minor21} study, who find a surprisingly steep density slope of $\eta\,=\,1.32\pm0.04$ from lensing alone. The paper reports a correlation between the derived density slope and the inferred subhalo mass; thus, a shallower lens slope would have implications for the claim of an unusually dense and massive halo. Their footnote 2 cites however a fairly modest decrease of $25\%$ in subhalo mass if the slope was confirmed to be approximately isothermal.

When drawing such comparisons, it is important to note the contrast in (and limitations of) the methods used in dynamical and pure lensing studies. In dynamical analyses, restrictive assumptions are often made on the lens galaxy axisymmetry and orbital structure; pure lensing studies typically assume simple power laws and a simplified linear external shear. As shown in this study, and also found in \citet{etherington23}, lensing-only studies appear to predict marginally steeper projected density slopes than lensing + dynamics studies do. If the observed discrepancies stem from a lack of complexity in the dynamical modelling, one would expect that as you relax the simplifications and introduce spatially resolved structure, the recovered slopes from the two methods should be in closer agreement. As we have shown, this is not the case, with our reported projected density slope showing less consistency with pure lensing slopes than that of the lensing + dynamics measurement from \citet{etherington23}'s one-aperture kinematics. Instead, we consider the possibility that the disparity arises from either deficiencies in the lens modelling, or more subtle limitations in the kinematics that can only be solved through more sophisticated dynamical modelling techniques such as Schwarzschild models \citep[][]{Schwarzschild79}. 

\subsection{Robustness of Assumptions}
\label{sec:dynamicallysimple}

As noted in Section~\ref{sec:modelling}, there exists a possible degeneracy between the models that maximise the likelihood in the inner and outer regions of the galaxy. There seems present an orthogonal angular dependence, such that we see a relatively high velocity dispersion along the minor axis at large radii, but conversely along the major axis at small radii. This is particularly evident in the velocity dispersion panel of Fig. \ref{fig:v_map}, where we suspect that this behaviour may be related to a greater influence from DM at this radii.

Given the apparent non-axisymmetry implied by the low surface-brightness envelope at large radii, and the possible signatures for past interactions \citep[as speculated by][]{Sonnenfeld12}, we excluded measurements from the nine Voronoi bins at ${r}\gtrsim2$\,arcsec from our preferred modelling.
If we instead fit to the full, unrestricted range of data, we recover a resulting projected density slope of $\eta\,=\,1.042\pm0.02$, which sits in slightly closer agreement with the lensing only studies, but the model now provides a poorer fit to the kinematics, especially in the outer regions.

The kinematics in non-axisymmetric mass distributions can in principle be tackled using more general dynamical models, such as the orbit-based approach of \citet{Schwarzschild79} \citep[e.g. see][]{Poci22}. This approach would perhaps mitigate the limitations imposed by our simple model, but would in turn demand much more stringent requirements on the data with a necessity for a very high signal-to-noise ratio that is unfeasible with the present observations. This is especially true when dealing with the pervasive contamination of source light at the first Einstein radius.

\subsection{Central Mass in Excess of a Constant Mass-to-Light Ratio}
\label{sec:BH?}

In the construction of our dynamical models we allowed for an additional compact mass, that is not described by the luminosity distribution or the NFW profile, and used this to describe any excess central mass. We find that the preferred excess central mass of our free $\gamma$ models is well constrained to be
$\sim8.23\times10^{9}$\,M$_{\odot}$, which if
attributed to a central black hole only, would be an outlier relative to black hole scaling relations \citep[e.g.][]{Gebhardt2000, Tremaine02, thomas16, VanDenBosch16}. Given the $M_{\text{BH}}-\sigma$ relation derived by \citet{VanDenBosch16}, it is expected that the true black hole contribution to the central mass component will be less than $\sim10^{9}$M$_{\odot}$. 
Although the form of the scaling relations at the highest masses is still uncertain \citep[e.g.][]{thomas16}, it is unlikely that the Jackpot lens galaxy truly harbours a $\sim10^{10}$\,$M_\odot$ black hole.

A more plausible explanation is that in our analysis we assume a constant $M/L_\star$ but, whilst we do not expect variations in age in galaxies of this type, there may be a metallicity gradient present, leading to a modest M/L$_\star$ gradient such as described by \citet{Tortora10}. 
Moreover, if there is a radial gradient in the stellar initial mass function as reported by \citet[][]{martin-navarro15, LaBarbera17, vandokkum17} \citep[but see][]{alton17, vaughan18}, then a much more substantial $M/L_\star$ gradient may be present, and indeed \citet{collett18} saw exactly this for a nearby lens. These works would suggest a steep increase in mass within $\lesssim$1\,kpc. Whilst we expect that the I-band image would be a faithful tracer of the stellar mass profile for a modest age and metalicity variation, this might not be true in the case of radial IMF variation; however, such a gradient should be absorbed into our central mass component, to first order, at the resolution of the present data.

In Appendix~\ref{sec:appendix} we show that tests with synthetic MUSE data suggest a potential for overestimation of the central mass. In the absence of a central mass in the input data, a mass comparable to that predicted by the standard $M_{\text{BH}}-\sigma$ relation was recovered. However, this is an order of magnitude smaller than the excess central mass recovered from the real data.

\section{Conclusions}
\label{sec:summary}

We have presented results from a kinematic and dynamical analysis of the Jackpot lens galaxy using new data obtained from a 5 hour MUSE integration, in order to constrain the 2D-projected total mass profile slope. To account for contamination from the source galaxy light, we implemented a multiple component fitting technique adapted from the {\sc ppxf} code that extracts the lens galaxy kinematics to first and second order. Simple gNFW + stars dynamical models were constructed with parameterised orbital anisotropies, DM density power-law slopes, stellar mass fractions and excess central mass components, and a robustly constrained aperture lensing mass was imposed. The posterior PDF for the model parameters was sampled, and a chi-squared likelihood maximised to derive the projected total density slope. 

This is the first 2D spatially resolved kinematics study for this system, and confirms the significant signature of rotation detected in previous studies. We measure rotation about the minor axis of $v\approx \pm 100$\,km\,s$^{-1}$ and a steep decrease in velocity dispersion from a central value of $\sigma \approx 290$\,km\,s$^{-1}$ to $\sigma \approx 200$\,km\,s$^{-1}$ in the outer regions. Notwithstanding the strong presence of rotation, the galaxy is dispersion-dominated at all radii. 
The kinematic measurements are consistent with those of previous single-slit studies of the Jackpot lens (i.e. central velocity dispersions of $\sigma_{\rmn{sonn12}}\,=\,287 \pm 11$\,km\,s$^{-1}$ and $\sigma_{\rmn{spin15}}\,=\,300 \pm 22$\,km\,s$^{-1}$), but are now fully mapped out in two-dimensions. 

From the {\sc jam} dynamical modelling, 
we infer a mass budget inside the Einstein radius that is dominated by stars ($\sim$70\,\%) if the halo slope is fixed to the NFW shape. For modified halos, 
we infer a larger DM fraction, and a DM density slope of $\gamma\,=\,1.73\substack{+0.17 \\ -0.26}$, which is significantly steeper than NFW (i.e. $\gamma\,=\,1$). This is in agreement with previous results for the Jackpot itself \citep[][]{Sonnenfeld12} and, at face value, supports the scenario in which DM haloes contract in response to the presence of a massive baryonic component. Indeed, our measured value of $\gamma$ is consistent with the gNFW slope of $\gamma\,=\,1.57$ used by \citep{sonnenfeld21} to model contracted haloes for massive lens galaxies. While similar conclusions have been reached by some ensemble studies of lens galaxies \citep[e.g.][]{grillo12}, others find 
that unmodified NFW haloes are preferred by the data \citep[e.g.][]{shajib21}. 

Our fitted models yield a 2D-projected total mass profile slope for the Jackpot lens of $1.03\pm0.03$, and a lensing-equivalent projected logarithmic density profile slope of $\eta\,=\,0.96\pm0.02$. Thus we confirm most-pure lensing results in finding a near isothermal profile \citep[e.g.][]{Collett14, etherington22}. Our profile is inconsistent with the surprisingly steep slope measurement of \citet{Minor21}.  

The main goal of this work, and of ongoing extensions relating to the Jackpot, is to 
suppress the remaining systematic errors and degeneracies, so as to fully exploit the cosmological potential offered by this unique lens system. An improved analysis of the lensing properties, exploiting multi-band imaging for all three sources, is presented by Ballard et al. (in preparation).
Future extensions to our work will incorporate the measured kinematics simultaneously with the lensing information, to create an even more detailed picture of the Jackpot system.

Additionally, having obtained spatially-resolved, high sensitivity, high resolution MUSE data for a larger sample of lens galaxies, we have begun to incorporate the techniques described in this paper to obtain robust ensemble total and DM density profile slope measurements. This will furthermore allow us to place improved constraints on the stellar initial mass function, the distribution of mass within galaxies and the structure of DM haloes, thus ultimately furthering our understanding of the intrinsic properties of galaxies and the nature of DM itself.

\section*{Acknowledgements}

We are grateful to Amy Etherington and James Nightingale for providing their parameterised source-subtracted model of Jackpot. This work is based on observations collected at the European Organisation for Astronomical Research in the Southern Hemisphere under ESO programme 102.A-0950. RJS was supported by the Science and Technology Facilities Council through the Durham Astronomy Consolidated Grants (ST/T000244/1 and ST/X001075/1). HCT acknowledges the support of the Science and Technology Facilities Council (STFC) studentship. TEC is funded by a Royal Society University Research Fellowship and the European Research Council (ERC) under the European Union’s Horizon 2020 research and innovation program (LensEra: grant agreement No. 945536).

\section*{Data Availability}

Data used in this paper are publicly available in the ESO and HST archives.


\bibliographystyle{mnras}
\bibliography{bib} 


\appendix

\section{Resolution and binning effects}
\label{sec:appendix}

Our dynamical analysis is performed using the standard {\sc jam} treatment commonly employed for $z\,\sim\,0$ galaxies, which typically have higher-resolution data than the data presented in this paper. We therefore use mock MUSE observations to assess the degree to which the resolution and binning of our observations affect the recovered parameters.

\begin{figure*}
	\includegraphics[width=1.8\columnwidth]{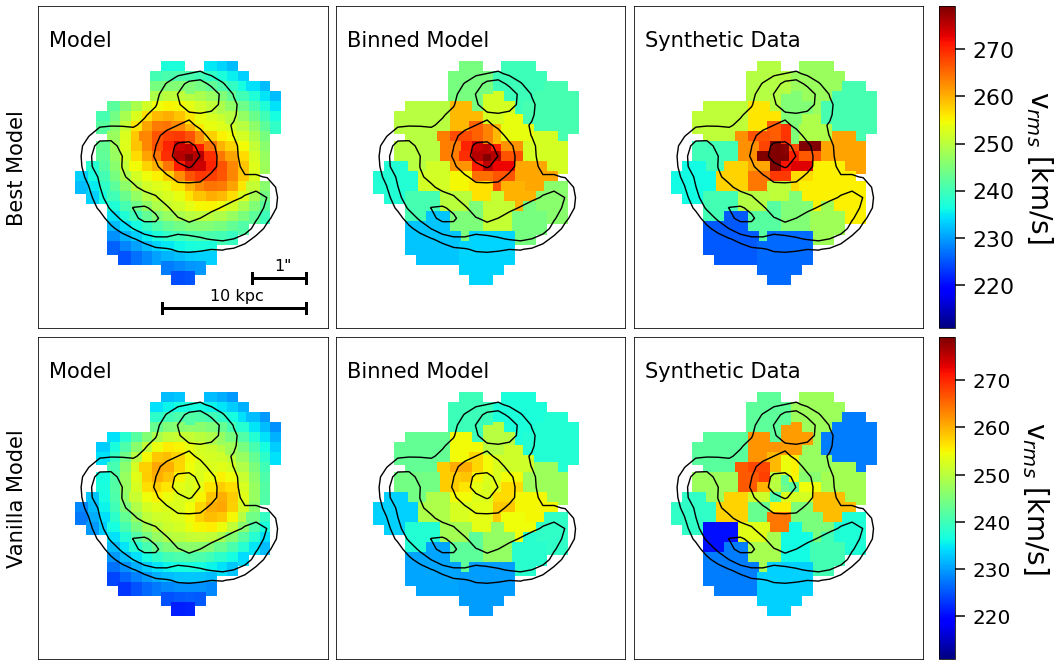}
    \caption{Kinematic maps showing the $v_{\text{rms}}$ fields of the synthetic data. We show two models, one created from the best-fitting free $\gamma$ model and one representing a `vanilla' scenario with no excess central mass and an NFW-like halo. The left hand panel shows the high resolution model, the middle panel is the noise-free spatially binned model and the third is the synthetic model after adding observational noise.
    }
    \label{fig:best_synth_map}
\end{figure*}

\begin{figure*}
	\includegraphics[width=1.8\columnwidth]{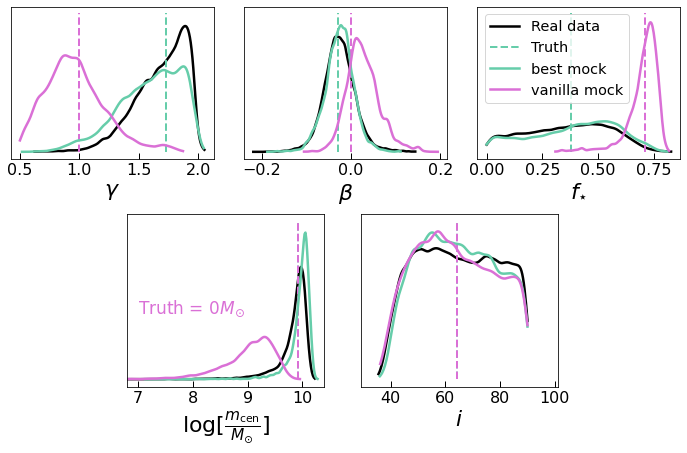}
    \hfill
    \caption{The marginalised posterior densities for each parameter for the real data and the two sets of mock data, generated from the `best' and `vanilla' models. The dashed line represents the `truth' value used to generate the mock data. The parameters explored are: the inner slope of the DM density profile, $\gamma$; the orbital anisotropy parameter, $\beta$; the stellar mass as a fraction of the total lensing mass, $f_{\star}$; any central mass in excess of a constant stellar mass-to-light ratio, $m_{\text{cen}}$; the galaxy inclination, $i$.}
    \label{fig:best_synth_corner}
\end{figure*}

We generate two synthetic data sets; the first set has properties similar to our best-fitting free $\gamma$ model from Section~\ref{sec:results}, and the second set represents an alternative case to test the recovery of parameters in a `vanilla' scenario with no excess central mass and an NFW-like DM halo. To generate these data sets, shown in Fig.~\ref{fig:best_synth_map}, we bin our best-fitting and vanilla models following the same Voronoi binning scheme described in Section~\ref{sec:kinematic_templates}. The binned $v_{\text{rms}}$ field of each model is convolved with the $v_{\text{rms}}$ error from our multiple component fitting from Section~\ref{sec:kin results}. The posterior PDF for the model parameters is sampled via a MCMC Ensemble sampler, following the techniques described in Section~\ref{sec:modelling}, and the likelihood of our mock data $v_{\text{rms}}$ is maximised. The marginalized parameter constraints and input `truth' parameters are shown in Fig.~\ref{fig:best_synth_corner} and summarised in Table~\ref{tab:table_app}.

\begin{table*}
	\centering
	\caption{The median and 68\% confidence bounds for the recovered model parameters from our mock data.}
    \def\arraystretch{1.25}%
	\begin{tabular}{ll|ccccc}
		\hline
		\multicolumn{2}{c|}{Model} & $\gamma$ & $\beta$ & $f_{\star}$ & $m_{\text{cen}}$[$10^{9}$M$_{\odot}$] & $i$[\textdegree]\\
		\hline
        Best & Input & $1.73$ & $-0.03$ & $0.38$ & $8.23$ & $64$\\
		 & Recovered & $1.63\substack{+0.24 \\ -0.31}$\ & $-0.02\substack{+0.03 \\ -0.03}$ & 
        $0.42\substack{+0.17 \\ -0.26}$ & $10.16\substack{+2.38 \\ -3.37}$ &
        $64\substack{+16 \\ -14}$\\
        \hline
         
        Vanilla & Input & $1$ & $0$ & $0.71$ & $0$ & $64$\\ 
         & Recovered & $0.94\substack{+0.27 \\ -0.22}$\ & $0.02\substack{+0.04 \\ -0.03}$ & 
        $0.72\substack{+0.04 \\ -0.06}$ & $1.31\substack{+1.46 \\ -0.92}$ &
        $62\substack{+17 \\ -14}$\\
        \hline

	\end{tabular}
    \label{tab:table_app}
\end{table*}

We find that the input parameters used to generate both models, with the exception of $m_{\text{cen}}$ for the `vanilla' mock data, are indeed recovered without bias from any smoothing effects, albeit with errors that are unavoidably larger than they would be for a more nearby galaxy. In the case of the `vanilla' mock data, in the absence of any real point mass or centrally concentrated mass in excess of a constant M/L$_\star$, the recovered excess central mass is of the order $10^{9}$M$_{\odot}$ and an order of magnitude smaller than what is recovered from the real data. This value falls below the threshold at which any real central excess mass could be confidently detected. 
Our synthetic data were generated from JAM predictions of the combined second moment, and therefore implicitly incorporate both the velocity and velocity dispersion. A more sophisticated treatment modelling both components consistently would be needed to address beam-smearing effects at the galaxy centre. However, given the relatively small contribution of the ordered rotation to the second moment for this specific galaxy, it is unlikely to have a significant impact on the results presented in this work. We thus conclude that the central mass recovered from the real data is not an artifact of smoothing or binning effects, but rather reflects the intrinsic properties of the system.

\bsp	
\label{lastpage}
\end{document}